\documentclass{aa} 
\usepackage{graphicx} 
\usepackage{txfonts} 
 
\usepackage{natbib} \bibpunct{(}{)}{;}{a}{}{,}

\def\HI{\ion{H}{i}}  \def\HeI{\ion{He}{i}} 
\def\HeII{\ion{He}{ii}}   
\def\CII{\ion{C}{ii}} \def\CIII{\ion{C}{iii}} \def\CIV{\ion{C}{iv}} 
   
 \def\NIV{\ion{N}{iv}}  
 \def\OII{\ion{O}{ii}} \def\OIII{\ion{O}{iii}} 
\def\OIV{\ion{O}{iv}} \def\OV{\ion{O}{v}}

\newcommand{\msunpyr}{\,M_\odot\,\mbox{yr}^{-1}}  
\newcommand{\kms}{\,\mbox{km}\,\mbox{s}^{-1}} 
\newcommand{\Hb}{\ifmmode {\rm H}\beta 
\else H$\beta$\fi}

\newcommand{\Op}{O$^{+}$} 
\newcommand{\Opp}{O$^{++}$}

\newcommand{\nii}{[N\,{\sc{ii}}]} 
\newcommand{\oiii}{[O\,{\sc{iii}}]} 
\newcommand{\sii}{[S\,{\sc{ii}}]} 
\newcommand{\ariv}{[Ar\,{\sc{iv}}]} 
 
\newcommand{\Oiitonea}{[O\,{\sc{ii}}]\,$\lambda 3726,3729$} 
\newcommand{\Oiitoneb}{[O\,{\sc{ii}}]\,$\lambda 7320,7330$}

\newcommand{\Oiiiuv}{O\,{\sc{iii}}]\,$\lambda 1661$} 
 
\newcommand{\Cii}{C\,{\sc{ii}}]\,$\lambda 2326$} 
\newcommand{\Ciii}{C\,{\sc{iii}}]\,$\lambda 1909$} 
\newcommand{\Ciir}{C\,{\sc{ii}}\,$\lambda 4267$} 
\newcommand{\Nii}{[N\,{\sc{ii}}]\,$\lambda 6584$} 
 
\newcommand{\Oi}{[O\,{\sc{i}}]\,$\lambda 6300$} 
\newcommand{\Oii}{[O\,{\sc{ii}}]\,$\lambda 3727$} 
\newcommand{\Oiii}{[O\,{\sc{iii}}]\,$\lambda 5007$} 
 
\newcommand{\Neiii}{[Ne\,{\sc{iii}}]\,$\lambda 3869$} 
 
\newcommand{\Siitonea}{[S\,{\sc{ii}}]\,$\lambda 6717,6731$} 
\newcommand{\Siitoneb}{[S\,{\sc{ii}}]\,$\lambda 4069,4076$} 
 
 \newcommand{\rNii}{[N\,{\sc{ii}}]\,$\lambda 5755 / 6584$} 
\newcommand{\rOiii}{[O\,{\sc{iii}}]\,$\lambda 4363 / 5007$} 
\newcommand{\rSii}{[S\,{\sc{ii}}]\,$\lambda 6731 / 6717$}

\newcommand{\cmcub}{~cm$^{-3}$} 

\newcommand{\msun}{\ifmmode M_{\odot} \else M$_{\odot}$\fi} 
\newcommand{\rsun}{\ifmmode R_{\odot} \else R$_{\odot}$\fi} 
\newcommand{\lsun}{\ifmmode L_{\odot} \else L$_{\odot}$\fi} 
\newcommand{\zsun}{\ifmmode Z_{\odot} \else $Z_{\odot}$\fi}

\newcommand{\Ha}{H$\alpha$}

\newcommand{\heii}{He\,{\sc ii}}

\newcommand{\oii}{[O\,{\sc ii}]} 
 
\newcommand{\neiii}{[Ne\,{\sc iii}]} 
 
\newcommand{\Sii}{[S\,{\sc ii}]$\lambda$6725} 
 
\newcommand{\Siiit}{[S\,{\sc iii}] $\lambda$6312} 
 
\newcommand{\Ariii}{[Ar\,{\sc iii}]$\lambda$7135} 
 
\newcommand{\Ariv}{[Ar\,{\sc iv}] $\lambda$\,$10^{-1}$4741}

\newcommand{\He}{He$^{0}$} 
 
\newcommand{\Cpp}{C$^{++}$}

\begin{document} 
 
\title{Comprehensive modelling of the planetary nebula LMC-SMP\,61 and its
  [WC]-type central star} \subtitle{}
  
\titlerunning{The planetary nebula LMC-SMP\,61 and its [WC]-type central star}
 
\author{G.\ Stasi\'nska\inst{1} \and G.\ Gr\"{a}fener\inst{2} \and M.\ 
  Pe\~{n}a\inst{3} \and W.-R.\ Hamann\inst{2} \and L.\ Koesterke\inst{4} \and
  R.\ Szczerba\inst{5}}
 
\institute{LUTH, Observatoire de Meudon, 5 Place Jules Janssen, F-92195 Meudon
  Cedex, France \\
  \email{grazyna.stasinska@obspm.fr} \and Institut f\"{u}r Physik, Astrophysik,
  Universit\"{a}t Potsdam, Am
  Neuen Palais 10, D-14469 Potsdam, Germany\\
  \email{goetz@astro.physik.uni-potsdam.de}
  \and Instituto de Astronom{\'\i}a, UNAM, Apdo. Postal 70 264, M\'exico D.F., 04510, M\'exico\\
  \email{miriam@astroscu.unam.mx}
  \and Laboratory for Astronomy and Solar Physics, NASA Goddard Space Flight Center, Code 681, Greenbelt, MD 20771, USA\\
  \email{lars@winds.gsfc.nasa.gov} \and N. Copernicus Astronomical Center,
  Rabia\'{n}ska 8,
  87--100 Toru\'{n}, Poland\\
  \email{szczerba@ncac.torun.pl}}
 
\offprints{G.\ Stasi\'nska}
 
\date{Received ; Accepted}
 
\abstract{ We present a comprehensive study of the Magellanic Cloud planetary
  nebula SMP\,61 and of its nucleus, a Wolf-Rayet type star classified [WC 5-6].
  The observational material consists of HST STIS spectroscopy and imaging,
  together with optical and UV spectroscopic data collected from the literature
  and infrared fluxes measured by IRAS.  We have performed a detailed spectral
  analysis of the central star, using the Potsdam code for expanding atmospheres
  in non-LTE.  For the central star we determine the following parameters:
  $L_\star$ = $10^{3.96} L_\odot$, $R_\star$ = $0.42\,R_\odot$, $T_\star$ =
  $87.5\,\mathrm{kK}$, $\dot{M}$ = $10^{-6.12}\msunpyr$, $v_\infty$ =
  $1400\kms$, and a clumping factor of $D$ = $4$. The elemental abundances by
  mass are $X_\mathrm{He}$ = $0.45$, $X_\mathrm{C}$ = $0.52$, $X_\mathrm{N} <
  5\,10^{-5}$, $X_\mathrm{O}$ = $0.03$, and $X_\mathrm{Fe}$ $< 1\,10^{-4}$.  The
  fluxes from the model stellar atmosphere were used to compute photoionization
  models of the nebula.  All the available observations, within their error
  bars, were used to constrain these models.  We find that the ionizing fluxes
  predicted by the stellar model are basically consistent with the fluxes needed
  by the photoionization model to reproduce the nebular emission, within the
  error margins. However, there are indications that the stellar model
  overestimates the number and hardness of Lyman continuum photons. The
  photoionization models imply a clumped density structure of the nebular
  material. The observed \Ciii$/$\Ciir\ line ratio implies the existence of
  carbon-rich clumps in the nebula. Such clumps are likely produced by stellar
  wind ejecta, possibly mixed with the nebular material. We discuss our results
  with regard to the stellar and nebular post-AGB evolution.  The observed
  Fe-deficiency for the central star indicates that the material which is now
  visible on the stellar surface has been exposed to s-process nucleosynthesis
  during previous thermal pulses.  The absence of nitrogen allows to set an
  upper limit to the remaining H-envelope mass after a possible AGB final
  thermal pulse.  Finally, we infer from the total amount of carbon detected in
  the nebula that the strong [WC] mass-loss may have been active only for a
  limited period during the post-AGB evolution.
  
   }
 
\maketitle

\section{Introduction} 
\label{sec:intro} 
 
Only a few studies have been devoted so far to a consistent modelling of a
planetary nebula and of its central star. Such studies are useful to get a
better insight into the relation between the nebula and its progenitor. Another,
very important aspect is that this is the only way to test model atmosphere
predictions in the Lyman continuum and thus to validate the model atmospheres.
The works of Rauch, K\"{o}ppen \& Werner (1994, 1996), Pe\~{n}a et al.  (1998),
De Marco \& Crowther (1998, 1999), De Marco et al.  (2001) are examples of such
studies, while Crowther et al. (1999) have performed a similar study on a
Population I Wolf-Rayet ring nebula.  Such investigations are particularly
important in the case of planetary nebulae with Wolf-Rayet type central stars
(which represent about 10\% of all planetary nebulae), since recent work (e.g.\ 
G\'orny \& Tylenda 2000, De Marco \& Soker 2002) have completely changed
previous views on the evolutionary status of these objects.

In general, Wolf-Rayet central stars of PNe belong to the [WC] sequence.  In our
galaxy most of these objects have been classified as [WC-early] or [WC-late]
types, with only few objects of intermediate types (Tylenda, Acker \& Stenholm
1993).  In the Magellanic Clouds, the WR central stars are also of [WC] type
(except for the extraordinary central star of LMC-N66, e.g.\ Pe\~na et al.
1997b), but in this case they have been classified in the intermediate [WC]
types.  Pe\~na et al.  (1997a) suggested that such a difference might be a
consequence of the differences in metallicity between the Milky Way and the
Magellanic Clouds.  In any case, [WC] central stars show spectral features
identical to those of massive WC stars but at much lower luminosity, and they
can be analyzed with the tools developed for massive WR stars (e.g.\ Hamann
1997).
 
In the present paper, we concentrate on the planetary nebula SMP\,61 (also known
as N203, WS 24 and LM1-37) which is among the brightest planetary nebulae in the
Large Magellanic Cloud. This is a good case for a detailed study: The object is
at a known distance modulus of 18.50\,mag \citep{ben1:02}, equivalent to
50.1\,kpc. The central star is one of the brightest [WC] central stars in the
LMC and therefore relatively easy to observe. It is of [WC\,5-6] type (Monk,
Barlow \& Clegg 1988; Pe\~na, Ruiz \& Torres-Peimbert 1997a), therefore the
nebular spectrum contains many lines of various excitation levels that allow
refined diagnostics. In addition, the nebula appears to be spherical and
integrated spectra are already available (Pe\~na et al.  1997a).

Using HST STIS, we have secured high signal-to-noise spectra of SMP\,61 in a
wide spectral range. This provided strong constraints for our modelling of the
central star atmosphere. The best fit model atmosphere was then used as an input
to build a photoionization model for the nebula.  The observations are described
in Sect.\,2, the stellar modelling technique is presented in Sect.\,3, the
spectral analysis of the central star of SMP\,61 is discussed in Sect.\,4. The
nebular modelling strategy is exposed in Sect.\,5 and the nebular model fitting
of SMP\,61 is presented in Sect.\,6.  The implications of our modelling are
discussed in Sect.\,7, and the main points of this study are summarized in
Sect.\,8.

\section{New Observations and Previous Data} 
\label{sec:obs} 
\subsection{Observations} 
HST STIS spectroscopic data were obtained on 1998/12/30 (HST Cycle 7, program ID
7303).  MAMA and CCD detectors were employed to cover the broadest spectral
range possible.  Prior to the spectroscopic observations, STIS images were
obtained to acquire the target.  The log of the observations is presented in
Table\,1.

\begin{table*}[htbp]
\caption{Log of observations } 
\smallskip 
\begin{tabular} {llcrcl} 
\hline 
 Instrument & Image ID & Date & $\lambda$ Range &  $t_{\rm exp}$ & Remarks 
 
                                                     \rule[-1mm]{0mm}{4.5mm}\\ 
           &       &          &  [\AA]     &  [min] &           
                                                     \rule[-2mm]{0mm}{4.5mm}\\ 
\hline 
 
 CTIO 4m Ret.2 $^{\rm a}$   & & 31/12/94 & 3184--7463  & 5.0 &   
                                          2 and 10\arcsec slit width \\ 
\hline 
 IUE SWP &  54313 & 06/04/95 & 1152--1978 & 135 & Large ap., low res.\\

\hline 
 HST STIS G140L&o57n02010&07/01/99&1124--1730&36.36 & FUV-MAMA \\ 
 HST STIS G140L&o57n02020&07/01/99&1124--1730&22 & FUV-MAMA \\ 
 HST STIS G230L&o57n02030&7/01/99&1580--3160&25.5 & NUV-MAMA \\ 
 HST STIS G230L&o57n02040&7/01/99&1580--3160&54 & NUV-MAMA \\ 
 HST STIS G430L&o57n03010&30/12/98&2900--5710&37.86 & CCD \\ 
 HST STIS image&o57n02iuq &06/01/99&5852&3.3 & CCD, acquisition im.\\ 
 HST STIS image&o57n03bqq &30/12/98&5852&3.3 & CCD, acquisition im.\\ 
\hline 
\end{tabular} 
 
(a) CTIO 4m with Reticon detector and grating KPGL2 \rule[0mm]{0mm}{4.5mm}\\ 
Also in HST archives FOS spectra Y2N30302T, Y2N30303T, Y2N30304T, 1150 -  4800 \AA \\ 
reported by Vassiliadis et al. (1998b) 
\label{table:observations} 
\end{table*} 
 
Direct images with HST WFPC were searched in the archives.  Apart from our own
STIS images for target acquisition (see Table\,1), we retrieved the FOC image
W1ID0601T, obtained on 1993/11/08, with exposure time of 3 min through the
filter F502N which isolated the strong \Oiii \ line (this image is shown in
Vassiliadis et al. 1998b).  We decided to use our target acquisition images to
analyze the nebular morphology and surface brightness because they are better
exposed and because the Vassiliadis et al.\ images were obtained before the
COSTAR mission. Our images are obtained with a filter including the brightest
emission lines: H$\alpha$, H$\beta$ and \oiii \ lines.
 
We also considered ground-based spectrophotometric data for SMP\,61 obtained by
Monk, Barlow, \& Clegg (1988), Meatheringham \& Dopita (1991) and Pe\~na et al.
(1997a).  Other available UV spectroscopic data are IUE data given in Pe\~na et
al. (1997) and HST FOS data from 1175 to 4800 \AA \ published by Vassiliadis et
al.  (1998a).
 
All the available spectroscopic data are presented in Table\,\ref{neb:fluxes},
which lists the reddening corrected fluxes for the observed lines (in units of
\Hb=100) obtained by the different authors. We also list the total H$\beta$ flux
observed in the slit, $F(\Hb)$, the logarithmic extinction at \Hb, $C(\Hb)$, and
the slit dimensions, when available. All the spectra mentioned above have good
signal-to-noise, and the differences among observations by different authors
are, in general, small.  Note that observations by Pe\~{na} et al. (1997a) were
made both with a slit of 2\arcsec and a slit of 10\arcsec and, as noted by these
authors, the line ratios are very similar in both slits. The \Hb\ flux reported
in column 2 of Table\,\ref{neb:fluxes} is the one obtained with the 10\arcsec
slit.
 
SMP\,61 is one of the twelve planetary nebulae in the LMC detected by IRAS. The
fluxes retrieved from the IRAS data base are: F(12$\mu$m) = 0.08 Jy (flux
density quality = 2) F(25$\mu$m) = 0.13 Jy (flux density quality = 3),
F(60$\mu$m)$\leq$0.16 Jy (flux density quality = 1).

 
\begin{table*}[hbt] 
\caption{Nebular dereddened fluxes from different authors } 
\smallskip 
\begin{tabular} {llrrrrr} 
\hline 
lambda & ion & \multicolumn{5}{c}{Dereddened fluxes relative to H$\beta$} 
\rule[-1mm]{0mm}{4.5mm}\\ 
\cline{3-7}         \rule[-1mm]{0mm}{4.5mm}                                             
    &        & STIS (this work) & PRT-P97a &  MD91  &Vass98a &MBC88           
                                              \rule[-2mm]{0mm}{4.5mm}\\
\hline 
  1240   &   N\,{\sc v}   & stell  &   ...  &         &   ...   &     \\ 
  1336   &  C\,{\sc ii}   &  ...   &   ...  &         &   26.0  &     \\ 
  1406   &  Si\,{\sc iv}],O\,{\sc iv}]& stell& ...  & &   12.9  &     \\ 
  1550   &  C\,{\sc iv}   &  stell &   stell&         &   72.1+50.4&    \\ 
  1640   &  \heii         &  stell &   ...  &         &   31.2  &    \\ 
  1649   &  C\,{\sc iv}   &  stell &   ...   &        &   10.3  &    \\ 
  1666   &  \oiii         &  6.8:  &    ...  &        &    7.4  &    \\ 
  1750   &  N\,{\sc iii}] &  0.09  &         &        &         &    \\ 
  1909   &  C\,{\sc iii}] & 316    &    401  &        &  392.7  &    \\ 
  2297   &  C\,{\sc iii}  &  stell &    ...  &        &   30.7  &    \\ 
  2321,26&  C\,{\sc ii}  &  62    &    ...  &        &   87.9  &    \\ 
  2405   &  C\,{\sc iv}   &  stell &    ...  &        &    8.1  &     \\ 
  2470   &  \oii          &  6.2:  &    ...  &        &    9.0  &    \\ 
  2524   &  C\,{\sc iv}   &  stell &    ...  &        &   11.9  &    \\ 
  3203   &  He\,{\sc i},\heii& stell&   ...  &        &    3.4  &     \\ 
  3726+29&  \oii          &    60  &   80.0  &   85.8 &         & $<$80  \\ 
  3835   &  H\,9          &   7.2  &    7.4  &   10.4 &         &   ...  \\ 
  3869   &  \neiii        &    43  &   41.6  &   50.6 &         &   47.2  \\ 
  3889   &  He\,{\sc i}+H\,8&  19  &  18.7   &   24.6 &         &    ...  \\ 
  3967+70&  \neiii+H\,7   &  28    &   29.0  &   37.9 &         &    ...  \\ 
  4026   &  He\,{\sc i}+\heii&  2.6&   2.5   &    3.9 &         &    ...  \\ 
  4069   &  \sii          &  2.2   &    2.9  &    3.9 &         &    ...  \\ 
  4102   & H$\delta$      &  26    &   25.2  &   31.2 &         &   13.5  \\ 
  4269   &  C\,{\sc ii}   &   1.8  &    1.4  &    1.7 &         &    ...  \\ 
  4340   & H$\gamma$      &  46    &   46.6  &   46.6 &         &   39.9  \\ 
  4363   & \oiii          &   5.3  &    5.7  &    6.4 &         &    5.5  \\ 
  4471   &  He\,{\sc i}   &   5.2  &    4.9  &    5.9 &         &    ...  \\ 
  4686   &  \heii         &   stell&    stell&    ... &         &    ...  \\ 
  4711+13&  \ariv+He\,{\sc i}& ... &     0.7 &    1.3 &         &    ...  \\ 
  4711   &  \ariv         &   ...  &    0.2: &        &         &         \\ 
  4740   &  \ariv         &   ...  &    0.2: &    ... &         &    ... \\ 
  4861   & H$\beta$       &  100.0 &   100.0 &  100.0 &         &  100.0 \\ 
  4921   &  He\,{\sc i}   &   1.5  &    1.4  &    2.0 &         &    ...  \\ 
  4959   &  \oiii         &   253  &  231    &  222.5 &         &   278.6 \\ 
  5007   &  \oiii         &   sat? &   sat   &   sat  &         &   800.0 \\ 
  5200   &  [N\,{\sc i}]  &        &    0.3: &    ... &         &    ...  \\ 
  5411   &  \heii         &        &    0.4  &    ... &         &    ...  \\ 
  5522   &  [Cl\,{\sc iii}] &      &    0.2: &    ... &         &    ...  \\ 
  5538   &  [Cl\,{\sc iii}] &      &    0.3: &    ... &         &    ...  \\ 
  5755   &  [N\,{\sc ii}]   &      &    0.7  &    1.4 &         &    ...  \\ 
  5808   &  C\,{\sc iv}     &      &   stell &    ... &         &    ...  \\ 
  5876   &  He\,{\sc i}     &      &   16.0  &  18.8  &         &   15.2  \\ 
  6300   &  [O\,{\sc i}]    &      &    4.3  &    6.1 &         &    3.3  \\ 
  6312   &  [S\,{\sc iii}]  &      &     1.2 &    1.3 &         &    ...  \\ 
  6363   &  [O\,{\sc i}]    &      &     1.3 &   2.3  &         &    ...  \\ 
   6548  &  \nii            &      &    15.8 &   15.0 &         &    7.7  \\ 
  6563   &  H$\alpha$       &      &   285.8 &  282.1 &         &  282.4  \\ 
  6584   &  \nii            &      &   31.5  &   44.8 &         &   22.3  \\ 
  6678   & He\,{\sc i}+\heii&      &    4.2  &    6.1 &         &    ...  \\ 
  6717   &  \sii            &      &    2.1  &    2.2 &         &    ...  \\ 
  6731   &  \sii            &      &    3.8  &   4.9  &         &    5.3  \\ 
  7065   &  He\,{\sc i}     &      &    8.2  &  13.7  &         &    ...  \\ 
  7136   &  [A\,{\sc iii}]  &      &    8.8  &   16.3 &         &    ...  \\ 
  7320+30& \oii             &      &   14.3  &   ...  &         &    ...  \\ 
\hline 
log $F(\Hb)$ &               & -12.80& -12.61 &  ...   &         &  ...    \\ 
$C(\Hb)$ &               & 0.12  &  0.18  &  0.19  &         & 1.07$^1$ \\ 
slit (\arcsec) &               & 0.2x52&  2;10  &  ...   &         &  3.6x3.6 \\ 
\hline 
\multicolumn{5}{l}{$^1$ value affected by atmospheric dispersion} 
\end{tabular} 
\label{neb:fluxes} 
\end{table*}

\subsection{Preliminary analysis of the observational data}

\begin{figure}[htbp] 
  \includegraphics[width=0.50\textwidth]{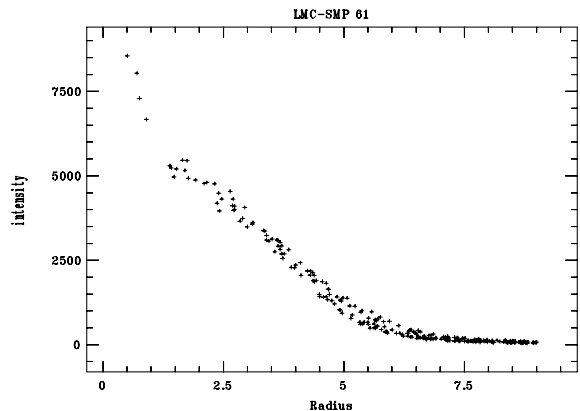}
\caption{
  Radial distribution of the surface brightness. The STIS image 57n02iuq,
  obtained for target acquisition, was employed to prepare this graphic. The
  radius is measured in pixels.  Each pixel is equivalent to 0.05\arcsec.}
  \label{fig:distribution} 
\end{figure}

In HST images the nebula appears perfectly spherical and shows no small-scale
structure.  The surface flux distribution, as derived from the image o57n02iuq,
is shown in Fig.\,\ref{fig:distribution}.  The ionizing star dominates the
emission in the central pixels. The nebula seems to have an inner radius of
about 0.065\arcsec, showing the maximum surface brightness there. The surface
brightness declines outwards and drops below 1/10 of its maximum at 0.26\arcsec.
 
The extinction of the nebula is modest. Throughout the paper, we use the value
$C(\Hb)$ = 0.18 (corresponding to $E_\mathrm{B-V} = 0.122\,\mathrm{mag}$ derived
by Pe\~na et al. (1997a) from the observed Balmer decrement and we adopt the
Seaton (1979) reddening law.
 
The STIS observations were essentially obtained to provide constraints on the
central star.  However, the 0.2\arcsec$\times$52\arcsec STIS slit includes a
significant fraction of the nebula so that both stellar and nebular emission is
seen. The nebular emission must of course be subtracted from the stellar
emission in order to perform the stellar analysis (see section 4.1).  Most of
the emission lines detected in the UV are of stellar origin, except for the
intense C\,{\sc iii}] 1909 \AA, C\,{\sc ii} 2321,26 \AA, and [O\,{\sc ii}] 2470
\AA. Many bright nebular lines were detected in the optical region.  The
dereddened fluxes of all the nebular lines measured in the STIS calibrated
spectra are presented in the first column of Table\,\ref{neb:fluxes}.
 
The STIS nebular lines together with the data presented by Pe\~na et al.
(1997a, column 2 in Table\,\ref{neb:fluxes}) have been used to derive first
estimates of the physical conditions and of the chemical composition of the
ionized gas. The atomic data are the same as used in the nebular modelling (see
Sect.\,5). The density derived from \rSii\ is 5400\cmcub, the temperature
derived from \rOiii\ is 10300\,K, and the one derived from \rNii\ is 11400\,K.
These values were used to derive the ionic abundances, and the ionization
correction factors from Kingsburgh \& Barlow (1994) led to the following total
abundance ratios : H : He : C : N : O : Ne : S : Ar = 1: .104 : $ 7.35\,10^{-4}$
: $3.95\,10^{-5}$ : $2.65\,10^{-4}$ : $4.26\,10^{-5}$ : $6.0\,10^{-6}$ :
$8.0\,10^{-7}$. These abundances are employed in Sect.\,6 as starting values for
the photoionization modelling.
 
Regarding the IRAS data, Zijlstra et al.  (1994) commented that the IRAS colors
F(12)/F(25)$\sim$0.62 and F(25)/F(60)$\leq$0.81 of SMP\,61 are very blue and
typical of a young nebula.

\section{Stellar modelling technique} 
\label{sec:stmod} 
 
\begin{figure*}[htbp] 
  \parbox[b]{0.65\textwidth}{\includegraphics[width=0.62\textwidth]{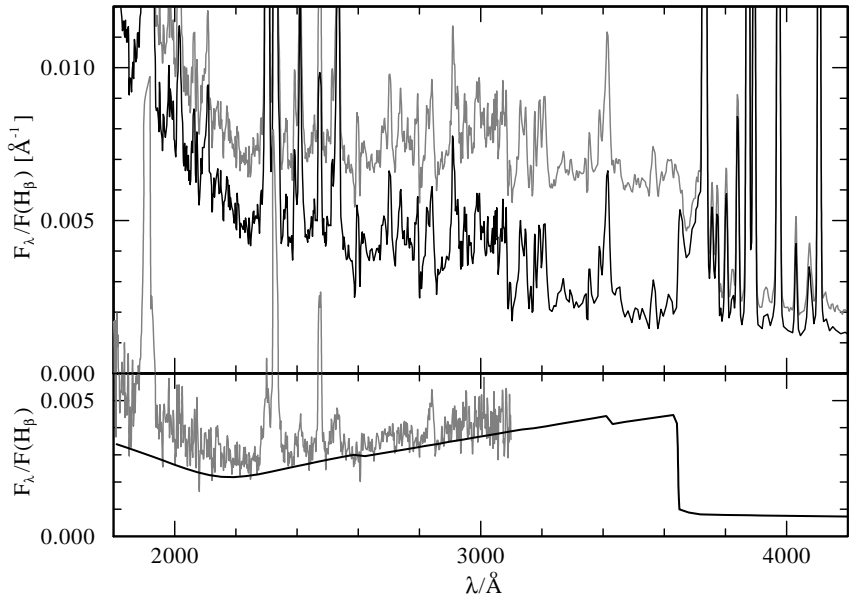}}
  \parbox[b]{0.32\textwidth}{\caption{ Subtraction of the nebular continuum.
      {\em Top panel}: The observed flux distribution around the Balmer jump
      before and after subtraction of the nebular continuum.  {\em Bottom
        panel}: The reddened theoretical continuum as utilized for the nebular
      subtraction is plotted in black.  It is compared to the nebular flux
      (grey) as estimated from the observation, by correcting for the stellar
      contribution (see text).  Although the latter may still include a small
      contribution from the central star, the theoretical flux distribution is
      clearly confirmed.}
      \label{fig:hstsub}} 
\end{figure*}

\subsection{Model atmospheres} 
\label{sec:atmos}

The analysis of hot stars with strong winds like the [WC]-type central star of
SMP\,61 requires the application of model atmospheres that account for complex
model atoms in non-LTE. Recent investigations revealed the necessity to include
iron line-blanketing as well as the effect of density inhomogeneities (clumping)
for a reliable determination of their stellar parameters and emergent flux
distributions \citep{ham1:98,hil1:99,cro1:02,gra1:02}.
 
In the present work, we utilize the Potsdam code for expanding atmospheres in
non-LTE \citep{koe1:92,ham1:92,leu1:94,koe1:95,leu1:96,ham1:98,koe1:02,gra1:02}.
This code calculates the radiative transfer in the co-moving frame of reference
for a spherically symmetric, stationary outflow. The atomic populations, the
electron density, and the electron temperature throughout the extended
atmosphere are determined from the equations of statistical- and radiative
equilibrium.  The system of statistical equations is solved in line with the
radiation transport by application of the ALI formalism \citep[accelerated
lambda iteration, see][]{ham2:85,ham1:86}, whereas radiative equilibrium is
obtained by a temperature correction procedure which is based on the method of
\citet{uns1:55} and \citet{luc1:64} \citep[see][ p.174]{ham1:03,mih1:78}.  The
statistical equations are solved for complex model atoms of He, C, O, Si, and
the iron group, taking advantage of the concept of super levels for the
inclusion of millions of iron-group line transitions \citep[see][]{gra1:02}.
Density inhomogeneities are accounted for by the assumption of small-scale
clumps with a constant volume filling factor $f_V$ \citep{ham1:98}.

\subsection{Model parameters} 
\label{sec:mpar} 
 
The model atmospheres are specified by the luminosity and radius of the stellar
core, by the chemical composition of the envelope, and its density- and velocity
structure. The basic parameters are: The stellar core radius $R_\star$ at
Rosseland optical depth $\tau_\mathrm{R} = 20$, the stellar temperature
$T_\star$ (related to the luminosity $L_\star$ via the Stefan Boltzmann law),
the chemical composition (given by mass fractions $X_{\rm He}$, $X_\mathrm{C}$,
$X_\mathrm{O}$, $X_\mathrm{Si}$, and $X_\mathrm{Fe}$), the mass-loss rate
$\dot{M}$, the terminal velocity $v_\infty$, and the ``clumping factor'' $D = 1
/ f_V$ giving the density enhancement of the clumped matter with respect to the
mean wind density. For the dependence of the velocity on radius, a relation of
the form $v(r)=v_\infty\;\left(1- R_0/r\right)$ is assumed.  This outer velocity
law is augmented by an exponentially decreasing density distribution in the
hydrostatic domain. $R_0$ is suitably determined to connect both domains
smoothly.

The parameters $R_\star$, $\dot{M}$, $v_\infty$, and $D$ are connected by the
``transformed radius'' $R_\mathrm{t} \propto R_\star ( v_\infty /
\sqrt{D}\dot{M} )^{2/3}$.  Models with the same $R_\mathrm{t}$ show almost
identical line equivalent widths \citep{sch1:89}. This invariance holds because
the line emission in WR stars is dominated by recombination processes
\citep[see][]{ham1:98}.

\section{Spectral analysis of the central star} 
\label{sec:spanal} 
 
In the present section, we perform a detailed spectral analysis of the central
star of SMP\,61 based on HST UV + optical observations. Owing to the reliable
spectrophotometry of HST in combination with the known distance and the moderate
interstellar reddening towards the LMC, the observations provide the absolute
stellar flux distribution in the wavelength range from 1150 to 6000\AA.  The
model comparisons in the present section are therefore performed with respect to
the absolute flux, corrected for interstellar extinction applying the standard
law of \citet{sea1:79} with $E_\mathrm{B-V} = 0.122\,\mathrm{mag}$.

\subsection{Subtraction of the nebular continuum} 
\label{sec:subtract} 
 
Due to the small angular diameter of the nebula ($0.7 \arcsec$),
the HST spectra ($0.2\arcsec$ slit width) include a significant part of the
total nebular flux.  Consequently, the observations clearly show nebular
features superposed to the stellar spectrum.  The most striking of these is the
Balmer jump at 3646\,\AA \ in emission, indicating the dominance of the nebular
continuum in the UV to blue part of the observation.  A reliable subtraction of
this continuum emission is essential for the spectral analysis of the central
star.
 
A distinction between both contributions is in principle possible, because the
nebula is spatially extended in contrast to the point-like star. To estimate the
contribution of the nebula, two spectra with apertures of $1.2\arcsec$ and
$0.25\arcsec$ are extracted from the UV part of the HST data.  The difference
between these two spectra is dominated by nebular emission. After scaling the
difference spectrum to the flux of the nebular \CII\ line at 2327\,\AA, a coarse
estimate for the pure nebular flux is obtained.
 
In Fig.\,\ref{fig:hstsub}, this spectrum is compared to the theoretical nebular
continuum flux computed with the physical conditions relevant to the object.
After applying the appropriate reddening the model flux is scaled by the ratio
of observed to theoretical $\mathrm{H}_\beta$ line flux.  Finally, the
theoretical continuum is utilized for the nebular subtraction.  Due to the good
agreement between theory and observation, a possible stellar contribution to the
observed Balmer jump can be excluded.

\subsection{Spectral fit} 
\label{sec:spfit}

\begin{figure*}[htbp] 
  \includegraphics[width=0.99\textwidth]{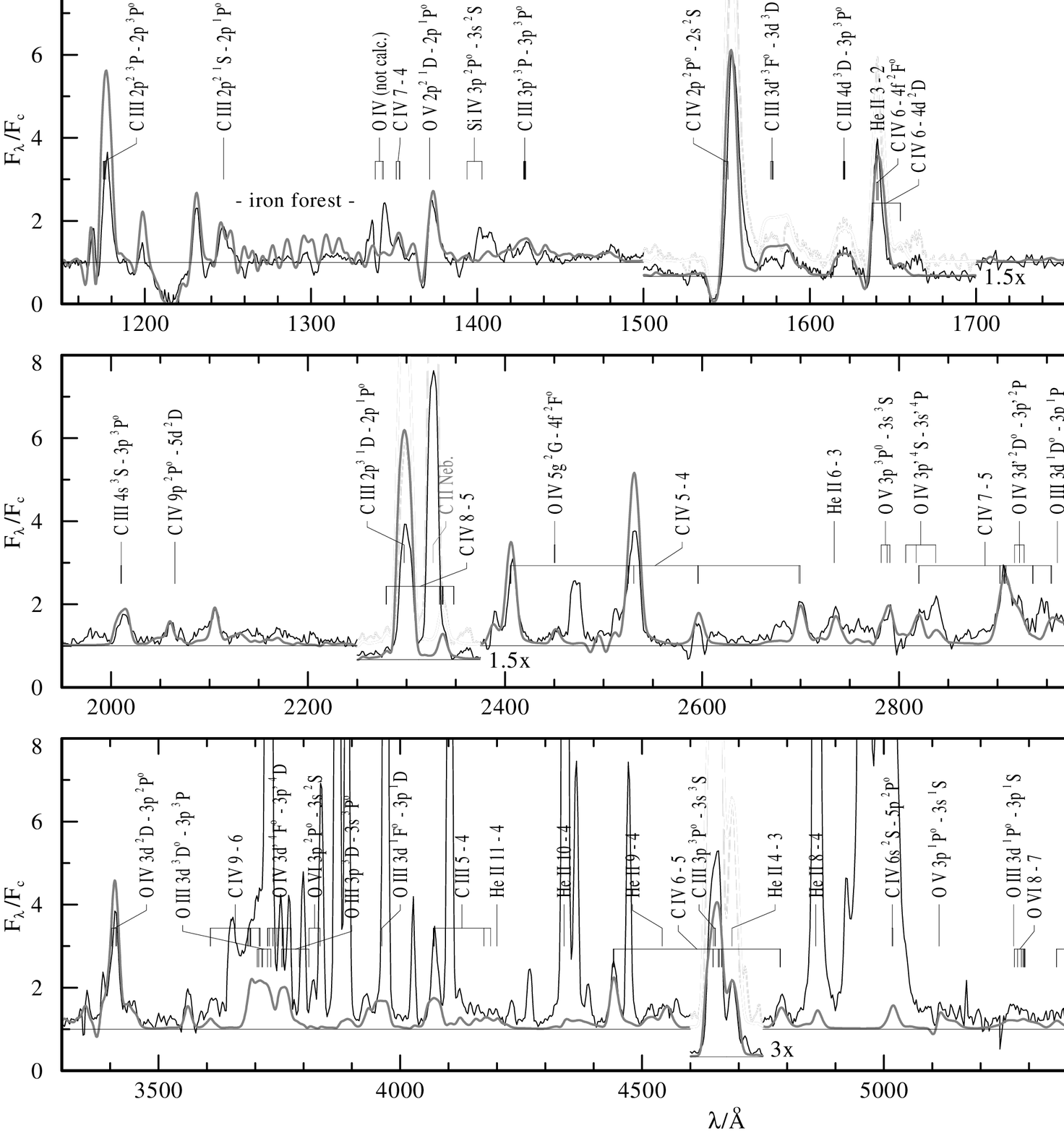}
  \caption{
    Spectral fit of the central star of SMP\,61.  The observation (thin line) is
    shown together with the synthetic spectrum (thick line, grey).  The model
    parameters are compiled in Table\,\ref{tab:parameters}.  Prominent spectral
    lines are identified.  After correction for interstellar extinction, the
    observed and the model flux are both divided by the {\em same} model
    continuum, i.e.\ absolute fluxes are compared. Additionally, a correction
    for interstellar Ly\,$\alpha$ absorption is applied to the model spectrum.}
  \label{fig:spec} 
\end{figure*} 
 
\begin{table}[tbp] 
  \caption{Model parameters for the central star of SMP\,61. 
 {\em Stellar core parameters}: 
 Stellar luminosity $L_\star$, 
 temperature $T_\star$ and radius $R_\star$. 
 {\em Wind parameters}: 
 mass-loss rate $\dot{M}$, 
 terminal wind velocity $v_\infty$, 
 clumping factor $D$, and the corresponding transformed radius $R_{\rm t}$. 
 {\em Atmospheric abundances}: 
 Mass fractions $X_{\rm He}$, $X_{\rm C}$, and $X_{\rm O}$ are determined by the model fit 
 (Fig.\,\ref{fig:spec}) whereas $X_{\rm Si}$ is set to a standard LMC value.  
 For $X_{\rm Fe}$ and $X_{\rm N}$ upper limits are given. 
 The hydrogen mass fraction $X_{\rm H}$ could not be determined in the present analysis. 
 {\em Interstellar parameters}: 
 Distance modulus $M - m$ for the LMC, 
 color excess due to interstellar extinction $E_{B-V}$, 
 and Ly\,$\alpha$ hydrogen column density $n_{\rm H}$.} 
  \centering 
  \begin{tabular}{llll} 
    \hline \hline 
    \rule{0cm}{2.2ex}$L_\star$ & $10^{3.96} L_\odot$\rule{2cm}{0cm}& 
    $R_\star$ & $0.42\,R_\odot$ \\ 
    $T_\star$ & $87.5\,\mathrm{kK}$ & & \\ 
    \hline 
    \rule{0cm}{2.2ex}$\dot{M}$ & $10^{-6.12}\msunpyr$ & $v_\infty$ & $1400\kms$ \\ 
    $D$ & $4$ & $R_\mathrm{t}$ & $4.63\,R_\odot$ \\ 
    \hline 
    \rule{0cm}{2.2ex}$X_\mathrm{H}$ & - & $X_\mathrm{He}$ & $0.45$ \\ 
    $X_\mathrm{C}$  & $0.52$  & $X_\mathrm{N}$  & $< 5\,10^{-5}$ \\ 
    $X_\mathrm{O}$  & $0.03$  & $X_\mathrm{Si}$ & ($2\,10^{-4}$) \\ 
    $X_\mathrm{Fe}$ & $< 1\,10^{-4}$ & & \\ 
    \hline 
    \rule{0cm}{2.2ex}$M - m$ & $18.5\,{\rm mag}$  & 
    $n_{\rm H}$ & $1\,10^{21}\,{\rm cm}^{-2}$ \\ 
    $E_{B-V}$ & $0.122\,\mathrm{mag}$ & & \\ 
    \hline 
  \end{tabular} 
  \label{tab:parameters} 
\end{table}

The central star of SMP\,61 shows a typical [WC\,5]-type spectrum dominated by
emission lines of He, C, and O.  Apart from a narrower line width, the spectral
appearance of SMP\,61 is similar to the well-known massive WC\,5 star WR\,111.
Consequently, the fit criteria and the fitting technique are similar to the
analysis of WR\,111 described by \citet{gra1:02}.  An important difference
between both objects concerns the low iron abundance of SMP\,61. The iron forest
around 1300\,\AA\ -- a combination of a large number of iron emission and
absorption lines -- is barely visible in the HST observations of this object.
 
The stellar analysis is based on fitting the absolute flux distribution
including the observed wind emission lines of \HeII, \CIII, \CIV, \OIII, \OIV,
and \OV. A hydrogen-free WC surface composition with a metallicity of $0.3$
solar is adopted (referring to the compilation of solar element abundances in
\citet[][p.\,318]{gra1:92}). This value matches the derived nebular oxygen
abundance.  The iron mass fraction is correspondingly set to
$X_\mathrm{Fe}=4\,10^{-4}$, and the silicon abundance to
$X_\mathrm{Si}=2\,10^{-4}$.
The other model parameters are kept free throughout the fitting procedure.  The
hydrogen abundance is set to zero. This assumption cannot be checked
empirically, because the strong nebular Balmer emission masks any possible
stellar contribution.  However, in previous works no hydrogen is found on the
surface of early-type [WC]\,stars, and even for late types only small amounts of
$X_\mathrm{H} < 0.1$ are detected \citep{koe1:01,dem2:01}.
 
The final model fit is presented in Fig.\,\ref{fig:spec} together with the
derived stellar parameters in Table\,\ref{tab:parameters}. As expected, the
stellar temperature, transformed radius, and surface composition are similar to
the values obtained for WR\,111 \citep{gra1:02}, while the luminosity is lower
by 1.5\,dex.  A good fit quality is obtained with some exceptions like
\CIII\,2297\,\AA\ (a transition that is mainly fed by dielectronic
recombination), the classification line \CIII\,5696\,\AA, or the \HeII/\CIV\ 
blend at 5412/5470\,\AA. The latter reacts very sensitively to changes of the
ratio of carbon to helium abundance.  Additionally, our model with standard iron
abundance produces too strong iron features in the UV.
 
\begin{figure}[htbp] 
  \includegraphics[width=0.47\textwidth]{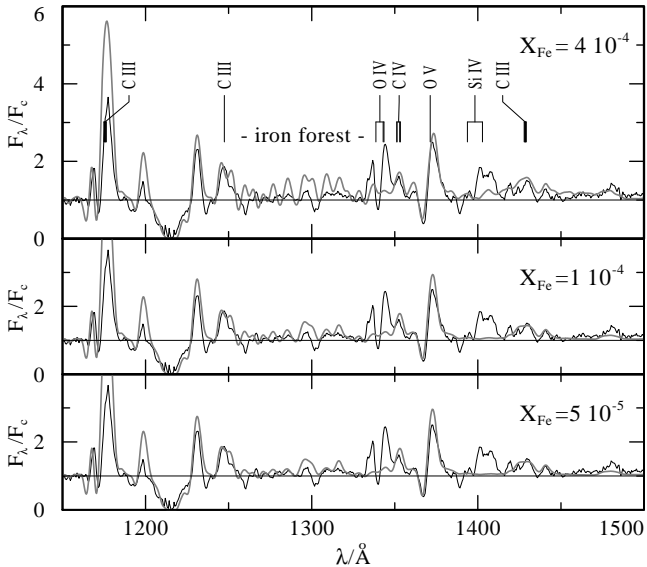}
  \caption{
    Model calculations for different iron abundances. The strength of the iron
    forest indicates a significant iron deficiency on the central stars surface
    ($X_\mathrm{Fe} < 1\,10^{-4}$).}
  \label{fig:iron} 
\end{figure} 
 
Test calculations for different iron abundances are presented in
Fig.\,\ref{fig:iron}. For a mass fraction around $5\,10^{-5}$ ($\approx 0.04$
solar) the overall strength of the iron forest is well reproduced, but the
wavelengths of the single transitions still do not match the observation.  For
this reason we can only set an upper limit of $X_\mathrm{Fe} < 1\,10^{-4}$
($\approx 0.07$ solar) to the iron abundance.  Compared to the standard LMC
metallicity found in the nebula the central star shows a significant iron
deficiency. In addition, the absence of \NIV\ line emission at 1721\,\AA, which
is relatively common among galactic [WC]\,stars, allows for the determination of
an upper limit for the nitrogen surface abundance. Calculations with a full
nitrogen model atom imply a value of $X_\mathrm{N} < 5\,10^{-5}$.
 
\begin{figure*}[htbp]   
  \parbox[b]{0.65\textwidth}{\includegraphics[width=0.62\textwidth]{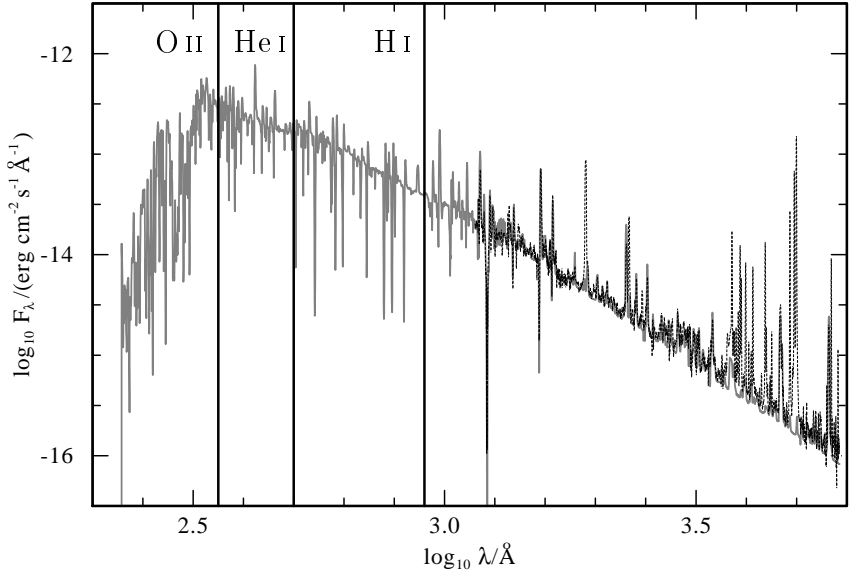}}
  \parbox[b]{0.32\textwidth}{\caption{ Overall flux distribution and ionizing
      fluxes of SMP\,61. The model flux is shown in grey together with the
      extinction-corrected observed stellar flux of SMP\,61 (black, dashed).
      The edge frequencies for the ionization of \HI, \HeI\ and \OII\ are
      indicated by vertical lines.}\label{fig:flux}}
\end{figure*} 
 
For the subsequent nebular analysis, it is important to discuss the reliability
of the derived ionizing fluxes.  The main difficulty in the present analysis
arises from the nebular continuum dominating the observed flux distribution from
$\approx 2500\mbox{\AA}$ to the visible wavelength range.  At first sight, the
situation seems to be well established, because the subtraction of the nebular
continuum works without difficulties. In addition, the resulting stellar flux
distribution is very well reproduced by our models, and the extinction parameter
determined from the nebular spectrum is perfectly compensating the interstellar
2200\,\AA \ feature (see Fig.\,\ref{fig:flux}). However, slight uncertainties of
the flux level in the optical wavelength range remain. For this reason, the
determination of the stellar luminosity primarily relies on the UV, i.e.\ it
becomes extremely sensitive to the applied reddening parameter $E_\mathrm{B-V}$.
 
Moreover, already small changes of the model parameters may significantly affect
the derived values for the stellar luminosity and ionizing fluxes, because the
observed flux distribution is still far away from the flux maximum at $\approx
350$\,\AA. Test calculations indicate that a variation of the model parameters,
leading towards an inferior but still acceptable fit quality, can change the
derived luminosity up to 0.1--0.2\,dex.  Including the observational
uncertainties mentioned above, the possible error may reach values up to
0.3\,dex.
 
In Fig.\,\ref{fig:flux}, the overall flux distribution as obtained from our
model calculations is shown together with the observed flux distribution, and
the ionization edges of \OII, \HeI\ and \HI.  The importance of testing the
ionizing fluxes for a reliable determination of the overall flux distribution of
the central star is obvious.  A detailed analysis of the nebular spectrum is
therefore capable to put substantial constraints on our atmosphere calculations.

\section{Nebular modelling technique}

\subsection{The photoionization code} 
 
The photoionization models for the nebula were constructed using the code PHOTO
which computes the radiative transfer of the ionizing photons in spherical
symmetry with the ``outward only'' assumption for the diffuse radiation field
and solves at each radius the equations of ionization equilibrium and thermal
balance.  The atomic data are the same as used by Stasi\'{n}ska \& Leitherer
(1996). The only modification is that low-temperature dielectronic recombination
coefficients for S and Ar ions were introduced (assuming that they are equal to
the computed values for the corresponding second-row elements, O and Ne
respectively). The effects of dust are treated as described in Stasi\'{n}ska \&
Szczerba (1999, 2001).
 
\subsection{The strategy} 
 
The aim is to reproduce all the observational constraints within a certain
degree of tolerance (to be clarified later). The observational constraints are
the emission line fluxes in the optical and the ultraviolet, the fluxes in the
12$\mu$m, 25$\mu$m and 60$\mu$m IRAS bands and the surface brightness
distribution as measured by HST. Out of the zone affected by the central star,
the surface brightness distribution is dominated by the
[O\,{\sc{iii}]\,$\lambda$4959, 5007}, \Hb\ and \Ha\ lines (with the \oiii\ lines
giving the major contribution except, perhaps, in the outer parts of the
nebula).
 
The models are computed using as an input the radiation field from the stellar
atmospheres described in the previous section.  The other input parameters are
the inner nebular radius, the gas density distribution, the abundances of the
elements and the characteristics of the dust grain population.
 
We assume that dust is composed of graphite grains with a classical power-law
size distribution of exponent -3.5 between 0.003$\mu$m and 0.3$\mu$m.  The total
dust-to-gas mass ratio is chosen so as to reproduce the observed flux in the
25$\mu$m IRAS band (which includes nebular fine structure lines). As a matter of
fact, in SMP\,61, the observed IRAS fluxes correspond to a small amount of dust,
which then does not affect significantly the nebular \Hb\ luminosity or its
ionization structure. We do not introduce an additional population of small
organic grains, as in Stasi\'nska \& Szczerba (2001), since the effects of those
grains are essentially to heat the gas and, as will be clear later, our initial
models rather show an excess of heating.

We first determine the density distribution that reproduces the observed surface
brightness distribution. By trial and error, we find that the observed surface
brightness profile is well reproduced by a spherical model with a hydrogen
density distribution described by
\begin {equation} 
  n = n_o \exp \left[ -\left(\frac{r-5\,10^{16}}{1.84\,10^{17}}\right)^{2}
  \right] ,
\end {equation} 
where $r$ is the distance to the star in cm and $n_{o}$ is the density at the
inner boundary.

The chemical composition with which we start is the one derived by classical
empirical methods in Sect.\,2.2. (the abundances of Mg, Si, Fe with respect to H
are fixed at an arbitrary value of $10^{-6}$ and do not affect our results).  In
a first step, we aim at reproducing the main characteristics of the nebula,
which can be summarized by the following quantities: F(\Hb) the total nebular
flux in \Hb, the nebular outer angular radius, $\theta$, defined as the radius
of the nebular image at zero intensity, and several line ratios: \rSii\ to test
the gas density, \rOiii\ to test the gas temperature in the \Opp\ zone,
\Oiitonea$/$\Oiitoneb, to test the density and temperature in the \Op\ zone;
\Oiii$/$\Oii\ to test the ionization structure; \Oiii$/$\Hb\ to test the heating
power of the central star, and \Oi$/$\Hb\ to test the importance of a warm
neutral zone. It is only when we are reasonably satisfied with these constraints
(which we will refer to as the restricted set) that we modify the elemental
abundances, if needed, to obtain an acceptable model.
 
In order to easily visualize whether a model is acceptable, we compute for each
constraint $P$ the quantity
\begin {equation} 
  \sigma_{\rm P} = ({\rm log} P_{\rm mod} - {\rm log} P_{\rm obs})/t,
\end {equation} 
where $P_{\rm mod}$ is the value returned by the model, $P_{\rm obs}$ is the
observed value, and $t$ the accepted tolerance in dex for this constraint. We
determine the values of $t$ for each constraint from the estimated uncertainties
in line fluxes, including the uncertainty in the dereddening process. Note that
we use the outer nebular radius $\theta$ as a convenient way to visualize the
constraint on the nebular size (but in fact we require the entire brightness
profile to be reproduced by the model). Rather than using line fluxes with
respect to \Hb, as is done usually, we prefer to use line ratios that are easier
to interpret. Apart from the traditional temperature and density diagnostics
mentioned above, we also use \Siitoneb$/$\Siitonea\ and \Oiiiuv$/$\Oiii, which
increase with nebular temperature and density, and \Ciii$/$\Ciir, which
increases with nebular temperature. The line ratios \Oiii$/$\Oii, \Ciii$/$\Cii,
N\,{\sc{iii}}]\,$\lambda$1750$/$\Nii,
[S\,{\sc{iii}}]\,$\lambda$6312$/$\Siitonea, and
[Ar\,{\sc{iv}}]\,$\lambda$4740$/$[Ar\,{\sc{iii}}]\,$\lambda$7136 reflect the
ionization structure of the nebula. Finally, \Nii$/$\Oii, \Sii$/$\Oii,
\Ciii$/$\Oiii, \Neiii$/$\Oiii\ and [Ar\,{\sc{iii}}]\,$\lambda$7136$/$\Oiii\ are
closely linked to the abundances of the elements with respect to oxygen while
the IRAS fluxes reflect the abundance and temperature of dust. The entire list
of constraints used in this study, together with the associated tolerances is
given in Table\,4. Note that we do not consider redundant lines, such as
[N\,{\sc{ii}}]\,$\lambda$6548 whose ratio with \Nii\ is independent of the
physical conditions in the nebula.  The results of our models are presented in
graphical form in the upcoming figures, where the values of $\sigma_{\rm P}$ are
plotted for all the constraints $P$. The quality of a model is then easily
judged by the location of the points with respect to the line of ordinate 0.
Note that we consider a model as fully satisfying only if \emph {each} of the
values of $\sigma_{\rm P}$ is roughly found between $-1$ and $+1$.  \emph {Any}
strong deviation requires an explanation.
 
\begin{table}[htbp] 
  \caption{Observational constraints and tolerances for the planetary nebula SMP\,61.} 
  \centering 
 
  \begin{tabular}{lll} 
    \hline \hline 
 
    \hline 
 
$F(\Hb)/10^{-13}$        & 3.71   &  0.18 \\ 
$\theta$               & 3.50\,$10^{-1}$   &  0.06 \\ 
He I 5876/H$\beta$          & 1.60\,$10^{-1}$   &  0.06 \\ 
\rSii\          & 1.81    &  0.11 \\ 
\rOiii\         & 7.15\,$10^{-3}$   &  0.08  \\ 
\rNii\         & 2.22\,$10^{-2}$   &  0.30 \\ 
\Oiitoneb/\Oiitonea\        & 1.79\,$10^{-1}$   &  0.08 \\ 
\Siitoneb/\Siitonea\   & 4.92\,$10^{-1}$   &  0.18 \\ 
\Oiiiuv/\Oiii    & 9.18\,$10^{-3}$   &  0.30 \\ 
\Ciii/\Ciir\      & 1.76$10^{2}$   &  0.15 \\ 
\Oiii/\Oii\   & 9.26    &  0.08 \\ 
\Ciii/\Cii\      & 5.10    &  0.11 \\ 
N\,{\sc{iii}}]\,$\lambda$1750/\Nii\     & 2.86\,$10^{-1}$   &  0.30 \\  
\Siiit/\Siitonea\   & 2.03\,$10^{-1}$   &  0.15 \\ 
\Ariv/\Ariii\  & 4.55\,$10^{-2}$   &  0.30 \\ 
\Oiii/H$\beta$        & 7.41    &  0.04 \\ 
\Oi/H$\beta$          & 4.20\,$10^{-2}$   &  0.15 \\ 
\Nii/\Oii\    & 3.94\,$10^{-1}$   &  0.08 \\ 
\Sii/\Oii\  & 7.37\,$10^{-2}$   &  0.15 \\ 
\Ciii/\Oiii\    & 4.26\,$10^{-1}$   &  0.11 \\ 
\Neiii/\Oiii\  & 5.67\,$10^{-2}$   &  0.06 \\ 
\Ariii/\Oiii\  & 1.19\,$10^{-2}$   &  0.10 \\ 
F(12$\mu$m)              & 8.00\,$10^{-2}$   &  0.18 \\ 
F(25$\mu$m)             & 1.30\,$10^{-1}$   &  0.10 \\ 
F(60$\mu$m)              & 1.60\,$10^{-1}$   &  0.18 \\ 
\hline

  \end{tabular} 
  \label{tab:constraints} 
\end{table}

\section{Results from the nebular model fitting} 
\label{sec:nebfit}

\subsection{First models} 
 
We use the stellar atmosphere model described above as an input to the
photoionization code. From the point of view of the nebular diagnostics
available, there is no difference between using the model atmosphere with iron
mass fraction $X_\mathrm{Fe}=4\,10^{-4}$ or $X_\mathrm{Fe}=5\,10^{-5}$.  The
hydrogen density distribution of the nebular model is given by Eq. (1).  As will
be made clear later on, one of the main shortcomings of the models is that the
excitation, as measured by \Oiii$/$\Oii\ is too high. We therefore discuss only
models tailored to provide the lowest possible \Oiii$/$\Oii\ ratio. For example,
there is no point to present models in which the nebula would have an inner
radius smaller than $5 \,10^{16}$\,cm (which corresponds to 0.065\arcsec at the
distance of the LMC). Also, while planetary nebulae are not necessarily
ionization bounded, optically thin models are not discussed, since they lead to
higher \Oiii$/$\Oii\ ratios than ionization-bounded models. We tried various
values of the inner density $n_{0}$. Note that the \rSii\ ratio is not very
sensitive to $n_{0}$, due to the fact that the hydrogen density decreases
outwards, and that, in addition, the electron density in the \Sii\ emitting zone
is smaller than the hydrogen density. We show in Fig.\,6 a series of models with
$n_{0}$ equal to $1.0 \,10^{4}$\cmcub\ (circles), $1.05 \,10^{4}$\cmcub\ 
(squares), $1.1 \,10^{4}$\cmcub\ (triangles), $1.15 \,10^{4}$\cmcub\ (diamonds).
All these models have a dust-to-gas mass ratio of $3\,10^{-4}$. We see that a
value of $n_{0}$ between about $1.02$ and $1.12\,10^{4}$\cmcub\ leads to an
angular radius (and a surface brightness profile) compatible with the
observations.  But they give too large $F(\Hb)$ (by a factor about 3 compared to
the observed value after correction for extinction). We cannot help by adding
more grains to absorb part of the Lyman continuum radiation, since the modelled
flux in the 25$\mu$m IRAS band is already larger than observed. Thus, we
conclude that, either the real number of stellar photons in the Lyman continuum
is smaller than in the model atmosphere or that there is some leakage of
ionizing photons. As seen in Fig.\,6, the photoionization models also predict
too large \Oiii$/$\Oii\ and \Oiii$/$\Hb\ (with a $\sigma_{\rm P}$ of 3--4), and
slightly too large \rOiii. This cannot be improved by changing the oxygen
abundance or by fine tuning the C/O and S/O ratios to fit the observed
\Ciii$/$\Oiii\ and \Sii$/$\Oii\ ratios. The models also have too low
\Oiitoneb$/$\Oiitonea\ and \Siitoneb$/$\Siitonea\ ratios, the agreement with the
observations being better for higher values of $n_{0}$, because the nebula is
then more efficient in absorbing photons and the ionization front occurs at a
higher value of the density.

\begin{figure}[tbp] 
  \includegraphics[width=0.50\textwidth]{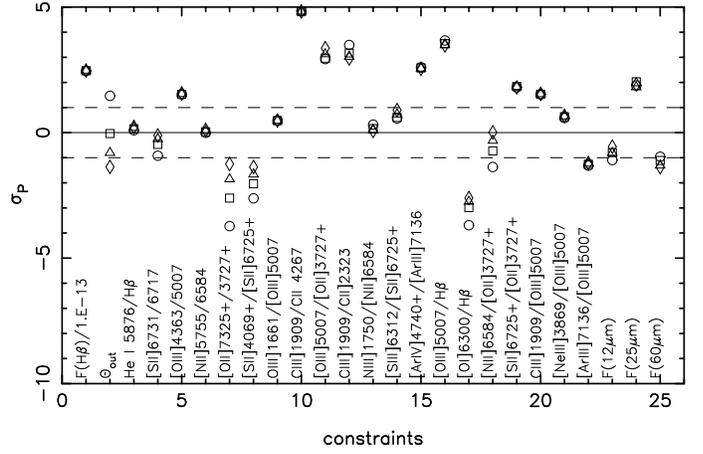}
\caption{
  Representation of a series of photoionization models of SMP\,61. The plotted
  quantities are the values of $\sigma_{\rm P} = ({\rm log} P_{\rm mod} - {\rm
    log} P_{\rm obs})/t$ as a function of the observational constraint $P$ (see
  text in Sect.\,5.2). The photoionization models have a density structure that
  reproduces the observed surface brightness profile and they use as an input
  the stellar atmosphere flux described in Sect.\,4. The chemical composition of
  these models is described in Sect.\,2.2. All the nebular models in this figure
  have a volume filling factor of one. Models with inner density
  $n_{0}$=$1.0\,10^{4}$, $1.05\,10^{4}$, $1.1\,10^{4}$, $1.15\,10^{4}$\cmcub\ 
  are represented by circles, squares, triangles and diamonds respectively.  In
  principle, a completely satisfactory model should have all its points located
  between the dashed lines. }
  \label{fig:6}  
\end{figure} 
 
In order to get away with the $F(\Hb)$ problem, one could argue that the
covering factor of the nebula is smaller than unity, so that some ionizing
photons escape without interacting with the gas. But this would not help in
solving the \Oiii$/$\Oii, \Oiii$/$\Hb\ and \rOiii\ problems.
 
We do not attempt two-sector models of different densities (such as those of
Clegg et al. 1987 or Dudziak et al. 2000) for the following reasons. First, as
mentioned in Sect.\,2, the nebular image is extremely round. This, a priori does
not mean that the nebula is spherical. Indeed, it could be elongated but seen
pole on. But, in that case, the nebular extension along the line of sight would
be larger than the one observed on the images. Replacing a portion of the model
nebula with two sectors of lower density (to obtain a larger dimension) would
considerably worsen the \Oiitoneb$/$\Oiitonea\ and \Siitoneb$/$\Siitonea\ ratios
while not helping much for the \Oiii$/$\Oii\ ratio and worsening the
\Oiii$/$\Hb\ ratio, as we can judge from spherical models built at lower
densities. Besides, if the low density sector is density bounded, this may
indeed help with the \Hb\ flux, but will worsen the \Oiii$/$\Oii\ ratio.

One way to lower the excitation of the nebula is to assume that it is clumpy.
Models that differ in density and filling factor but have the same value of the
product $n_{0}^{2} \epsilon$, where $\epsilon$ is the volume filling factor,
will have about the same external radius but will have lower excitation for
higher densities. Such a series of models is shown in Fig.\,7, where the values
of the couple ($n_{0}$, $\epsilon$) are ($1.05\,10^{4}$\cmcub, 1.0) (circles),
($1.5\,10^{4}$\cmcub, 0.49) (squares), ($2.0\,10^{4}$\cmcub, 0.27) (triangles),
($3.0\,10^{4}$\cmcub, 0.122) (diamonds). As seen in the figure, models with
higher densities considerably improve many of the line ratios, and especially
the \Oiii$/$\Oii\ and \Oiii$/$\Hb\ ratios as well as the \Oiitoneb$/$\Oiitonea\ 
and \Siitoneb$/$\Siitonea\ ratios and also the \Oi$/$\Hb\ ratio.  It is then
easy to fine tune the N, S and C abundances. A model with $n_{0}$ between
$2.0\,10^{4}$ and $3.0\,10^{4}$\cmcub\ would then be in agreement with all the
observational constraints (except \Ciii$/$\Ciir, but see later) if one assumed a
covering factor of roughly 1/3.
 
\begin{figure}[tbp] 
  \includegraphics[width=0.50\textwidth]{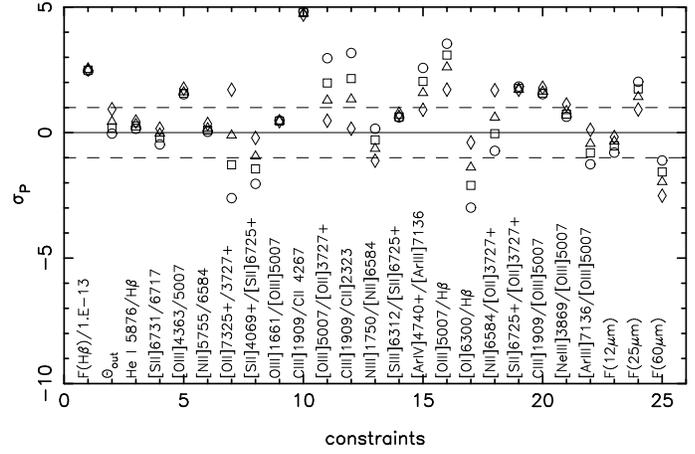}
\caption{
  Same as Fig.\,6 but for models with different filling factors. Models defined
  by ($n_{0}$, $\epsilon$) equal to ($1.05\,10^{4}$\cmcub, 1.0) are represented
  by circles, ($1.5\,10^{4}$\cmcub, 0.49) by squares, ($2.0\,10^{4}$\cmcub, 0.27
  by triangles, ($3.0\,10^{4}$\cmcub, 0.122) by diamonds.}
  \label{fig:7}  
\end{figure}

The predicted emission in the IRAS bands deserves some comments. First, we note
in Fig.\,7 that models with lower excitation show lower fluxes in the IRAS
bands. The reason is that the larger proportion of neutral hydrogen particles in
the nebula reduces the amount of Lyman continuum photons absorbed by dust.
Another point to make is that, in the hypothesis of incomplete coverage of the
ionizing source by the nebula (and if dust is intimately mixed with the gas),
the predicted IRAS fluxes should be larger than observed by roughly the same
amount as $F(\Hb)$ is larger than observed in our models computed for a covering
factor of unity. Obviously, this is not the case of our predicted flux in the
12$\mu$m band (the observed flux in the 60$\mu$m band is only an upper limit so
there is no problem there). Increasing the dust abundance only worsens the
problem, since then the dust particles are more numerous to share the photons
that heat them, and the temperature of the dust grains becomes smaller,
decreasing the flux in the 12$\mu$m band with respect to that in the 25$\mu$m
band. Actually, there may be a simple explanation to our under-prediction of the
flux in the 12$\mu$m band. Our code does not handle PAHs, while PAHs are known
to exist in planetary nebulae with Wolf-Rayet central stars (Szczerba et al.
2001, Cohen 2001) and they contribute to the emission in the 12$\mu$m band. Note
also that our predictions concerning dust emission are given here only for the
ionized part of the nebula. The presence of dust in the neutral zone would
increase the fluxes in the $25\mu$m and $60\mu$m bands. We do not consider this
in the modelling presented here, since it brings no useful information in the
problem we are interested in. In the following, we therefore simply require from
our models that they do not over-predict the 12$\mu$m and 25$\mu$m fluxes with
respect to the observations.
 
We have found that with a model having an inner density of about
$2-3\,10^{4}$\cmcub\ and a filling factor about $0.3-0.1$, we are able to
reproduce many of the observational constraints, provided that we assume a
covering factor of about 1/3. However, this model is still not entirely
satisfactory. Mainly, the \oiii\ temperature is still too high and so is the
predicted \Oiii$/$\Hb\ ratio. And a conspicuous shortcoming of the model is that
the \Ciir$/$\Ciii\ ratio is predicted too large by a factor of about 5. Note
that, while the density structure of planetary nebulae is known generally not to
be uniform on a small scale from high resolution images, the clumps or filaments
of gas are actually embedded in a more diffuse medium. This diffuse medium is
expected to be more highly ionized, and thus its presence will enhance the
\Oiii$/$\Oii\ predicted by our simple model. We do not attempt to construct a
two-phase model but we keep this in mind.
 
In the following, we look for more satisfactory solutions by releasing some of
the assumptions of our models.

\subsection{Models with modified stellar fluxes} 
 
Due to the compact and spherical appearance of the nebula, the covering factor
of 1/3 which is needed to explain $F(\Hb)$ seems to be very unrealistic.  The
obvious next step is therefore to assume that the discrepancy is at least partly
due to an overestimation of the stellar ionizing radiation field in
Sect.\,\ref{sec:spanal}. As explained in Sect.\,\ref{sec:spfit}, the error for
the derived stellar luminosity may be as high as a factor of 2 under very
pessimistic assumptions.

Let us therefore first consider the case where we assume that the number of
photons emitted by the star at any energy larger than 13.6\,eV is only 50\% of
what is predicted by the model atmosphere (from the point of view of the
photoionization model, it is equivalent to assume that the star luminosity is
lower than given in Sect.\,4 by a factor 2). The results are shown in Fig.\,8.
Here, the values of the couple $n_{0}$, $\epsilon$ are $7.5\,10^{3}$\cmcub, 1.0
(circles), $1.0\,10^{4}$\cmcub, 0.6 (squares), $1.2\,10^{4}$\cmcub, 0.4
(triangles), $1.4\,10^{4}$\cmcub, 0.3 (diamonds) while the dust-to-gas mass
ratio is still assumed to be $3\,10^{-4}$. Since the number of ionizing photons
is smaller than in the previous case, we had to take smaller densities to match
the observed size of the ionized nebula. Globally, the ionization parameter is
nevertheless smaller, so that \Oiii$/$\Oii\ can now be fitted even with a model
with a filling factor of unity. Of course, the \Oiitoneb$/$\Oiitonea\ and
\Siitoneb$/$\Siitonea\ ratios are better reproduced for the highest density
models of this series, i.e. $n_{0}$ around $1.5\,10^{4}$\cmcub, which again
require a small filling factor. The best models of this series are at least as
good as the best models shown in Fig.\,7 without requiring a small covering
factor.  Note that the \rOiii\ ratio is still somewhat high in this series of
models, and adjusting the S, C and Ne abundances (by lowering them) can only
make the problem worse.
 
\begin{figure}[tbp] 
  \includegraphics[width=0.50\textwidth]{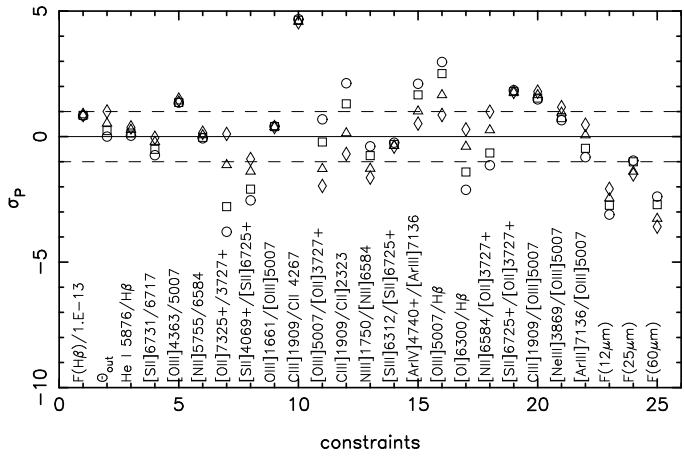}
\caption{
  Same as Fig.\,7 but for models with in which the stellar photon flux has been
  divided by a factor 2 with respect to the original model for all the energies
  larger than 13.6\,eV.  Models defined by ($n_{0}$, $\epsilon$) equal to
  ($7.5\,10^{3}$\cmcub, 1.0) are represented by circles, ($1\,10^{4}$\cmcub,
  0.6) by squares, ($1.2\,10^{4}$\cmcub, 0.4) by triangles, ($1.4,10^{4}$\cmcub,
  0.3) by diamonds.  }
  \label{fig:8}  
\end{figure} 
 
Another option to consider is a radiation field softened in the Lyman continuum.
This would not only lower the \Oiii$/$\Oii\ ratio, but also reduce the heating.
Fig.\,9 shows a series of models identical to those of Fig.\,8, but this time
keeping 60\% of the photons between 13.6 and 24.6\,eV and 40\% of the photons
with energies larger than 24.6\,eV. Clearly, the models from this series
represented by the triangles are quite satisfactory also as regards the electron
temperature and the \Oiii$/$\Hb\ ratio. \Sii$/$\Oii\ can easily be matched by
reducing the original sulfur abundance by a factor $1.3 - 2$). The observed
fluxes in the IRAS 25\,$\mu$m and 60\,$\mu$m bands are compatible with
dust-to-gas mass ratios of up to $3\,10^{-3}$. The quality of the fit of the
optical and UV data is not affected by the amount of dust in the allowed range.
The only important mismatch of the model with the observations is the
\Ciii$/$\Ciir\ whose computed value is much higher than observed, by a factor of
almost 4. This problem is discussed in the following subsection.

\begin{figure}[tbp] 
  \includegraphics[width=0.50\textwidth]{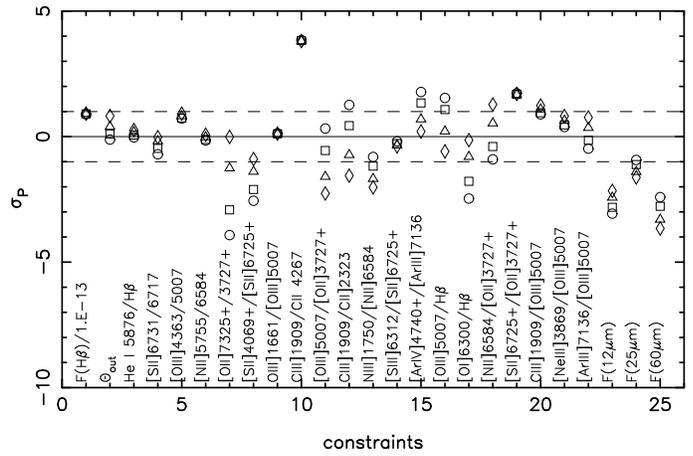}
  \caption{
    The same as Fig.\,8 for models ionized by a softened stellar flux (see
    text).  Models defined by ($n_{0}$, $\epsilon$) equal to
    ($7.5\,10^{3}$\cmcub, 1.0) are represented by circles, ($110^{4}$\cmcub,
    0.6) by squares, ($1.2\,10^{4}$\cmcub, 0.4 by triangles,
    ($1.4,10^{4}$\cmcub, 0.3) by diamonds.  }
  \label{fig:9}  
\end{figure}

\subsection{Models with inhomogeneous nebular composition} 
\label{sec:neb_composition} 
It is a known fact that recombination lines of C, and also of other elements
like N, O, Ne generally indicate much higher abundances of the parent ions than
collisionally excited lines (see e.g.\ Liu 2002 or Esteban 2002 and references
therein). Many explanations have been proposed (see e.g.\ Stasi\'nska 2002 for a
review). The generally preferred one is to assume that the chemical composition
of the nebulae is not uniform (e.g.\ Torres-Peimbert, Peimbert \& Pe\~{n}a 1990,
Liu et al. 2000).  And indeed, there is \emph{ direct} evidence at least in two
nebulae (A\,30 and A\,78) of the presence of clumps of matter enriched in C
(Jacoby \& Ford 1983, Harrington \& Feibelman 1984) and also N, O and Ne for A30
(Wesson, Liu \& Barlow 2003). The central star of SMP\,61 being a Wolf-Rayet,
this nebula is a good candidate for carbon enrichment. If material seen in the
stellar wind ends up in the nebula as cool clumps, these clumps will strongly
emit in the C\,{\sc{ii}}\,$\lambda$4267 line. We have mimicked such a situation
by assuming that the inner zone of the nebula consists of helium- and
carbon-rich material, maintaining the same density distribution as before. The
number of combinations to explore is obviously very high, and we have run many
simulations. In Fig.\,10, we present the result of a series of simulations with
$n_{0}$=$1.4\,10^{4}$\cmcub, $\epsilon$=0.3 (the best combination from Fig.\,9)
with the following abundances of He and C (relative to H by number) in the inner
3\,10$^{-3}$\msun\ of the nebula : 1.04 and 0.4 (circles), 0.78 and 0.3
(squares), 0.52 and 0.2 (triangles), 0.26 and 0.1 (diamonds). These ratios
correspond to the mass ratio of He and C in the atmosphere of the central star,
as derived in Sect.\,4. Otherwise, the number abundance ratios with respect to
hydrogen are: H : He : C : N : O : Ne : S : Ar = 1 : .09 : $ 4.0\,10^{-4}$ :
$3.95\,10^{-5}$ : $2.65\,10^{-4}$ : $4.26\,10^{-5}$ : $4.0\,10^{-6}$ :
$8\,10^{-7}$. We see that the last combination provides a very good fit to the
data, including the C\,{\sc{ii}}]\,$\lambda$1909$/$C\,{\sc{ii}}\,$\lambda$4267
line ratio. Probably, other combinations can be found that provide a similar
fit.

\begin{figure}[tbp] 
  \includegraphics[width=0.50\textwidth]{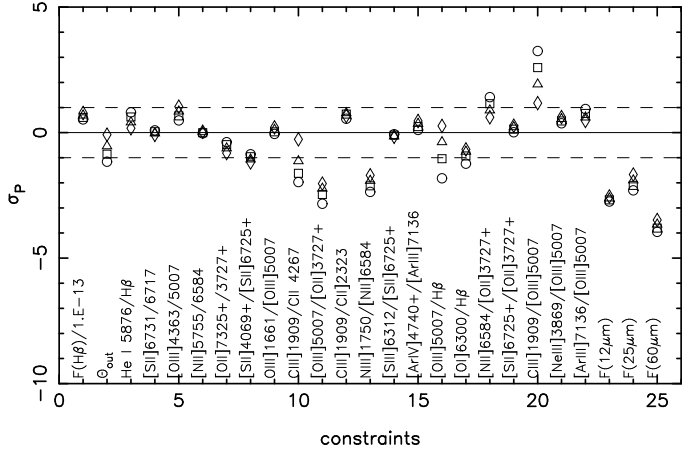}
  \caption{
    Models ionized by a softened stellar flux (as in Fig.\,9), with an inner
    density $n_{0}$=$1.4\,10^{4}$\cmcub and a nebular filling factor
    $\epsilon$=0.3 (the best combination of Fig.\,9). Now, the inner
    $3\,10^{-3}$\,\msun\ of the nebula are He- and C-rich. The abundances of He
    and C (by number relative to H) are: 1.04 and 0.4 (circles), 0.78 and 0.3
    (squares), 0.52 and 0.2 (triangles), 0.26 and 0.1 (diamonds).  Otherwise,
    the abundance ratios with respect to hydrogen are (by number): H : He : C :
    N : O : Ne : S : Ar = 1: .09: $ 4.0\,10^{-4}$ : $3.95\,10^{-5}$ :
    $2.65\,10^{-4}$ : $4.26\,10^{-5}$ : $4.0\,10^{-6}$ : $8\,10^{-7}$. }
  \label{fig:10}  
\end{figure} 
 
\section{Discussion}

\subsection{Stellar ionizing fluxes} 
 
The nebular analysis of SMP\,61 indicates two possible problems concerning the
theoretical flux distribution of the central star: The absolute number of Lyman
photons is probably too large by a factor of 2, and the Lyman continuum
radiation field is possibly too hard. The first problem may be discussed away by
assuming a very pessimistic value for the uncertainty of the derived luminosity.
However, the second point concerns the slope of the energy distribution in the
flux maximum and cannot be easily resolved.
 
A simultaneous solution of both problems would be achieved by decreasing the
effective temperature $T_\mathrm{eff}$ of the model atmosphere (i.e. increasing
the radius where $\tau_\mathrm{Ross} = 2/3$). In this way, photons would be
distributed from the flux maximum to the observed wavelength range. In effect,
the derived luminosity would be decreased, and a softer flux distribution would
be obtained in the flux maximum. For a corresponding decrease of
$T_\mathrm{eff}$, the stellar core temperature must be lowered significantly, or
the mass-loss rate must be increased. Of course, both operations are strongly
limited by the necessity to fit the observed spectrum of the central star. Test
calculations show that the simultaneous fitting of the observed spectrum {\em
  and} the nebular ionizing fluxes is not possible with the model atmospheres
applied in present work.
 
A possible solution concerns the treatment of dielectronic recombination in our
stellar atmosphere code. In the present version dielectronic transitions are
treated as optically thin, and their contribution to the rate equations is
accounted for by the approach of \citet{mih1:71}.  However, when \CIV\ 
recombines to \CIII\ in the outer part of the WC atmosphere, the ionization edge
of \CIII\ becomes optically thick.  In this case also the corresponding
dielectronic transitions become optically thick, and their recombination
efficiency is reduced because recombination photons cannot escape.  Indeed,
first tests with an improved treatment show that the recombination from \CIV\ to
\CIII\ is suppressed by this effect, leading to higher derived mass-loss rates
and slightly lower stellar temperatures. A detailed investigation of this topic
will be presented in a forthcoming paper.

\subsection{The effect of He-rich clumps on the nebular spectrum } 
 
He-rich clumps, which we have invoked in Sect.\,6.3, can in fact soften the
radiation field available to the rest of the nebula under certain conditions, by
selectively absorbing the photons above 24.6\,eV.  The total mass of helium
$M_{\rm He}^{\rm a}$ required to absorb all the \He-ionizing photons emitted by
the central star is of the order $10^{3}$/$n_{\rm e}$\,\msun\ (where $n_{\rm e}$
is the electron density in \cmcub). If the integrated mass of helium in the
clumps is at least equal to this value and if the clumps are located close to
the inner boundary of the nebula with an integrated covering factor of unity,
then the \He-ionizing photons would be completely blocked by the He-rich
material. Such an extreme situation does not correspond to the case of SMP\,61,
since \Oiii\ is emitted not only in the most central part but in the entire
nebula.  One can however imagine a less extreme situation where the high density
He- and C-rich clumps are distributed over the entire volume of the nebula and
soften the average radiation field available to the bulk of the nebular
material. In such a case, a softening of the predicted stellar energy
distribution in the Lyman continuum would perhaps not be necessary. Note however
that such clumps are not efficient in absorbing radiation between 13.6 and
24.6\,eV, unless they are very dusty, which for SMP\,61 seems excluded by the
observed IRAS fluxes. Obviously, the impact of these He- and C- rich clumps on
the global spectrum of the nebula will strongly depend on their density, on the
amount of mixing with nebular material and on their spatial distribution in the
nebula.

In the case of SMP\,61, we can estimate the maximum mass of helium contained in
the clumps by considering that the He/C proportion is the same as in the stellar
wind and by using the observed flux in the \Ciir\ line. The observed flux in
this line corresponds to a carbon mass $M_{\rm C}$ of about $10$/$n_{\rm
  e}$\,\msun, if most of the carbon is in \Cpp\ form and if the line is emitted
at a temperature of 8000\,K, which is the case for the C-rich zone in our
models. Since helium and carbon are roughly equal by mass in the stellar wind,
this means that the total helium mass is much smaller than to $M_{\rm He}^{\rm
  a}$, implying that there is no significant blocking of \He-ionizing photons by
the clumps. If the clumps are of much higher density than in our models, \Cpp\ 
may be partly recombined in the clumps, increasing $M_{\rm C}$ with respect to
$M_{\rm He}^{\rm a}$, but in that case the clumps will have a smaller cross
section and will be less efficient to block the stellar radiation.  Therefore,
in the specific case of SMP\,61, we do not think that the He-rich material will
significantly soften the stellar radiation available to the nebula.

\subsection{Stellar parameters and evolutionary status} 
 
The derived stellar parameters are in line with recent analyses of galactic
[WC]-type central stars based on line-blanketed models
\citep{dem1:98,dem1:99,dem1:01,cro1:03}. In contrast to previous un-blanketed
calculations \citep{koe1:97,koe2:97}, the new models show a trend towards
similar surface mass fractions for early {\em and} late [WC]\,subtypes, with
$X_\mathrm{O}\approx10\%$ and $X_\mathrm{C}/X_\mathrm{He}\approx1$.
 
Due to the known distance to the LMC, SMP\,61 offers the rare chance to
determine the luminosity of a [WC]-type central star. From the analysis of the
stellar spectrum alone we infer a value of $10^{3.9}$\lsun. The nebular analysis
reveals that this value is probably too large by up to 0.3\,dex -- dependent on
the assumed nebular covering factor. The luminosity of SMP\,61 is therefore in
the range of $10^{3.6}$--$10^{3.9}$\lsun.  For a second object, BD+30\,3696,
\citet{lij1:02} derive a distance of 1.2\,kpc from the angular expansion of the
nebula. For this distance, \citet{cro1:03} obtain a value of $L_\star =
10^{3.6}L_\odot$. Both values are in the range that is expected for the majority
of all post-AGB stars, with a typical mass around 0.6\,\msun\ 
\citep[see][]{blo1:95,her2:01}. Altogether, no evidence is found that
[WC]\,central stars have different masses than H-rich objects although,
admittedly, the number of cases with known $L_\star$ is small.
 
An interesting hint on the evolutionary status of SMP\,61 is given by its low
iron and nitrogen abundances. For a standard solar composition as given by
\citet[][p.\,318]{gra1:92} or \citet{gre1:98} a ratio of
$\log(\mathrm{Fe}/\mathrm{O})=-1.33$ is expected. More recent investigations
\citep{all2:01,all1:01} imply a lower solar oxygen abundance and give
$\log(\mathrm{Fe}/\mathrm{O})=-1.19$. The upper limit of $X_\mathrm{Fe} <
1\,10^{-4}$ that is derived in Sect.\,\ref{sec:spanal} translates to
$\log(\mathrm{Fe}/\mathrm{H}) < -5.54$ for a solar composition. In relation to
the oxygen abundance of the nebula ($\log(\mathrm{O}/\mathrm{H}) = -3.58$) we
obtain $\log(\mathrm{Fe}/\mathrm{O}) < -1.96$, i.e.\ iron is under-abundant by
at least 0.63\,dex in the atmosphere of the central star. In addition, no
nitrogen is detected in our analysis. The upper limit of $X_\mathrm{N} <
5\,10^{-5}$ lies 0.9\,dex below the abundance derived for the nebula.

The detection of an iron deficiency is in line with recent results from analyses
of PG\,1159 stars, the probable descendants of [WC]-type central stars.
\citet{mik1:02} set upper limits from 0 to -1.5\,dex solar for the iron
abundance of 15 PG\,1159 stars.  Additionally, \citet{wer1:03} find an iron
depletion of at least -1.5\,dex for the central star of A78, a [WC]-PG\,1159
transition object. Also \citet{cro1:03} find evidence for an iron abundance of
0.3--0.7\,dex below solar for BD+30\,3696 and NGC\,40. However, for these two
objects the nebulae show oxygen abundances significantly below solar
\citep[see][]{pen1:01}, so that the iron deficiency may be attributed to a low
initial metallicity.
 
Such an iron deficiency is expected for material which has been exposed to
s-process nucleosynthesis in the He-intershell of thermally pulsating AGB or
post-AGB stars \citep[see][]{lug1:03,her1:03}. This region is expected to
consist of partially He-burned material, i.e.\ mainly He, C, O and {\em no} N.
The observed surface composition of SMP\,61 therefore exactly resembles the
abundance pattern that is expected for a He-intershell where the s-process has
been active.

For the formation of central stars with [WC] surface composition, the H-rich
layers above the He-intershell must be removed or mixed with intershell
material. \citet{her3:01} and \citet{blo1:01} demonstrated that the latter is
possible by dredge-up of intershell material after an AGB-final thermal pulse
(AFTP) or a late- or very late thermal pulse on the post-AGB (LTP, VLTP). In
case of a LTP or VLTP very long evolutionary timescales on the post-AGB are
expected, because the central star first evolves to the blue as H-burner, then
back to the red during the thermal pulse, and again to the blue as He-burning
[WC]\,star.

The observed absence of nitrogen for SMP\,61 strongly favors the AFTP scenario
because for a LTP or VLTP nitrogen is produced by the CNO-cycle in the H-burning
phase. In the AFTP scenario the last thermal pulse occurs when the star is
leaving the AGB, and the central star directly enters the post-AGB as a
He-burner.  However, to obtain a sufficiently high probability for the AFTP to
occur at small enough H-envelope masses, a coupling of mass-loss to the thermal
pulse cycle is required \citep{blo1:01}.  From the upper limit of $X_\mathrm{N}
< 5\,10^{-5}$ we can derive an upper limit for the mass of the remaining
H-envelope after the final thermal pulse.

Under the assumption that nitrogen has nebular abundance in the H-envelope, and
is completely destroyed in the He-intershell, it follows that the maximum mass
of the H-envelope can be only 0.15 of the dredged-up mass from the intershell.
\citet{her3:01} finds that the dredged-up mass is typically of the order of
$6\,10^{-3}$\msun. Consequently, the remaining H-envelope mass at the final
thermal pulse must have been smaller than $9\,10^{-4}$\msun. This value is
significantly smaller than the corresponding envelope masses in the AFTP models
of \citet{her3:01}. For these models the mass-loss on the AGB has been tuned in
such a way that envelope masses of $3\,10^{-2}$ and $4\,10^{-3}$\msun\ were
obtained after the thermal pulse. The present work implies even lower values,
and therefore also lower H-abundances for the resulting [WC]-type central stars.

Additional evidence against the LTP and VLTP scenarios is due to the very young
appearance of the planetary nebula. Using the observed nebular expansion
velocity of $29.3\kms$ \citep{vas1:98} and angular radius from Table\,4 we find
an expansion time of about 3000\,yr, whereas evolutionary tracks of
\citet{blo1:95,blo1:01} for a 0.625\,\msun\ central star undergoing a VLTP imply
a timescale of at least 8000\,yr to reach a temperature of $85\,\mathrm{kK}$
(dependent on the time where the zero point for the PN age is assumed).
Moreover, \citet{gor1:00} find general evidence against the LTP and VLTP
scenarios due to the similarity of the [WR]PN and non-[WR]PN populations.

\subsection{Stellar mass-loss history} 
 
We know from the analysis of the central star atmosphere that the presently
observed mass-loss rate is $\dot{M}$ = $10^{-6.12}\msunpyr$ and that the carbon
mass fraction is 0.52 (see Table\,3). As mentioned in
Sect.\,\ref{sec:neb_composition}, the total mass of carbon in the nebula
required to produce the observed flux in the \Ciir\ line, $M_{\rm C}$, is about
$10$/$n_{\rm e}$\,\msun. In our model, the value of $n_{\rm e}$ in the zone
where this line is mostly emitted is about $2\,10^{4} $\cmcub\ which implies
$M_{\rm C}$ $\approx 5\,10^{-4}\msun$. Note that $M_{\rm C}$ is just slightly
larger than the total mass of carbon in the C-rich zone, and that, if the
density of the carbon-rich clumps is larger, the total carbon mass will be
smaller. We thus infer that the star has spent about $1300$\,yr in a similar
state of mass-loss.
 
This timescale is significantly lower than the time expected for a PN central
star to reach $T_\mathrm{eff} = 85\,\mathrm{kK}$. Evolutionary tracks for
He-burning stars of the corresponding luminosity from \citet{vas1:94} give
values around 3000\,yr.  \citet{vas1:98} show that this timescale fits the
nebular age of SMP\,61.  H-burning tracks from \citet{blo1:95} imply an age of
3000\,yr for a post-AGB mass of 0.625\,\msun\ (with $L_\star = 10^{3.9}\lsun$)
and 5000\,yr for 0.605\,\msun\ (with $L_\star = 10^{3.7}\lsun$).
 
Our modelling then suggests that the mass-loss of SMP\,61 may be intermittent,
which would indicate that it is initiated by other processes than radiative
acceleration. Interestingly, our atmosphere models would confirm this
hypothesis, because the force due to radiation pressure is much too low to
explain the observed mass loss: For the massive WC\,star WR\,111, which has a
very similar spectral appearance to SMP\,61, the radiation pressure, as
calculated in our models, supplies about one half of the energy necessary to
drive the stellar wind \citep{gra1:02}. On the other hand, for SMP\,61 the same
models provide only 17\% of the wind energy.  However, one must be aware that
the timescales for mass-loss and evolution derived above both suffer from large
uncertainties.  The spectroscopically derived mass-loss rate depends on the
clumping factor ($\dot{M} \propto 1/\sqrt{D}$ for constant $R_\mathrm{t}$, see
Sect.\,\ref{sec:mpar}) which is only roughly known.  The error margin for $\dot
M$ may therefore be as large as $\pm50\%$.  The stellar evolutionary timescale
is even more uncertain, because it depends on the stellar luminosity and the
mass of the stellar envelope above the burning shell. The latter is strongly
dependent on model assumptions. The nebular age is also uncertain, mostly
because the nebular velocity changes during the course of evolution as shown by
dynamical simulations (Mellema 1994, Villaver et al. 2002).

\section{Summary} 
 
We have obtained HST STIS imaging and spectroscopic observations of the
Magellanic Cloud planetary nebula SMP\,61 and its central star. We have also
searched the literature for other observational data concerning this object, in
order to obtain a full set of observational constrains and to assess their
uncertainties.
 
We have performed a detailed spectral analysis of the central star. For that, we
used the Potsdam code for expanding atmospheres in non-LTE, which includes the
statistical equations for many ions, including ions from the iron group.  The
observed stellar spectrum is well represented by a model with the following
parameters: $L_\star$ = $10^{3.96} L_\odot$, $R_\star$ = $0.42\,R_\odot$,
$T_\star$ = $87.5\,\mathrm{kK}$, $\dot{M}$ = $10^{-6.12}\msunpyr$, $v_\infty$ =
$1400\kms$, clumping factor $D$ = $4$ and $R_\mathrm{t}$ = $4.63\,R_\odot$. The
abundances of the elements by mass are $X_\mathrm{He}$ = $0.45$, $X_\mathrm{C}$
= $0.52$, $X_\mathrm{N} < 5\,10^{-5}$, $X_\mathrm{O}$ = $0.03$, $X_\mathrm{Fe}$
$< 1\,10^{-4}$. The low iron and nitrogen abundances indicate that the star has
undergone an AGB final thermal pulse with a very low H-envelope mass ($<
9\,10^{-4} \msun$). The iron depletion is probably due to s-process
nucleosynthesis during previous thermal pulses.

The fluxes from the model stellar atmosphere were used as an input to our
photoionization code to construct photoionization models of the nebula.  We
considered all the available observational constraints, within their error bars.
These constraints are numerous and allow a detailed discussion of the models.
We find that the fluxes from the stellar atmospheres allow to reproduce many
features of the nebular emission. However, the observed nebular properties are
better explained if the central star produces less photons in the Lyman
continuum than the model atmosphere by about a factor 2. An even better fit is
obtained if the Lyman continuum radiation is also softer than predicted by the
stellar model.  A possible solution involves the treatment of dielectronic
recombination in the stellar atmosphere code and will be presented in a
forthcoming paper.
 
We also find that the nebula must contain extremely carbon rich clumps in order
to reproduce the observed \Ciii$/$\Ciir\ ratio. The observational constraints do
not allow us to describe these clumps in detail (density, chemical composition,
location) but these clumps are likely produced by ejecta from the clumpy stellar
wind, possibly partly mixed with the nebular material. Obviously, high signal-to
noise and high spatial resolution observations of nearby planetary nebulae
excited by a similar star would be extremely useful to better understand the
process of enrichment of planetary nebulae by freshly made carbon (and helium).

From a comparison of the estimated total amount of carbon in the nebula with the
carbon mass-loss rate measured in the central star, we infer that the strong WC
mass-loss may have been active for only a limited period in the post-AGB
evolution of SMP\,61. However, this last result relies on uncertain timescales
and is subject to revision.

\begin{acknowledgements} 
 
  This work has been partially supported by DGAPA/UNAM (grant 114601),
  CONACYT/M\'exico (grant 32594-E), by the CNRS-CONACYT joint project (number
  10385)
  and by the Deutsche Agentur f\"ur Raumfahrtangelegenheiten under grant
  \mbox{DARA 50\,OR\,0008}.
 
\end{acknowledgements}



\end{document}